\documentclass[12pt,fleqn]{iopart}
\usepackage{iopams}
\usepackage{amssymb,mathptm,latexsym}

\def\be{\begin{equation}}
\def\ee{\end{equation}}
\def\bea{\begin{eqnarray}}
\def\eea{\end{eqnarray}}
\def\ba{\begin{array}}
\def\ea{\end{array}}
\def\nn{\nonumber}
\def\p{\partial}
\def\cD{\mathcal{D}}
\def\cL{\mathcal{L}}

\begin{document}


\title[Symmetry operators and separability of massive Klein-Gordon and Dirac equations
...]{Symmetry operators and separability of massive Klein-Gordon and Dirac equations in
the general 5-dimensional Kerr-(anti-)de Sitter black hole background}
\author{Shuang-Qing Wu}
\address{College of Physical Science and Technology, Central China
Normal University, Wuhan, Hubei 430079, People's Republic of China \\
Institute of Particle Physics, Hua-Zhong Normal University, Wuhan, Hubei 430079,
People's Republic of China} 
\ead{sqwu@phy.ccnu.edu.cn}

\date{today}

\begin{abstract}
It is shown that the Dirac equation is separable by variables in a five-dimensional rotating
Kerr-(anti-)de Sitter black hole with two independent angular momenta. A first order symmetry
operator that commutes with the Dirac operator is constructed in terms of a rank-three
Killing-Yano tensor whose square is a second order symmetric St\"{a}ckel-Killing tensor
admitted by the five-dimensional Kerr-(anti-)de Sitter spacetime. We highlight the construction
procedure of such a symmetry operator. In addition, the first law of black hole thermodynamics
has been extended to the case that the cosmological constant can be viewed as a thermodynamical
variable.

\end{abstract}

\submitto{\CQG}


\maketitle


\section{Introduction}

It is well known that the Kerr metric \cite{RK} possesses a number of miraculous properties
called by Chandrasekhar \cite{SC1}, such as the separability of the geodesic Hamilton-Jacobi
equation \cite{BC1}, separability of a Klein-Gordon scalar field equation \cite{BC1},
separability and decoupling of the massless nonzero-spin field equations \cite{TME}. These
properties are later shown to be closely connected with the existence of a quadratic integral
of motion --- Carter's fourth constant that is associated with a second order symmetric
St\"{a}ckel-Killing tensor discovered in the Kerr metric by Carter \cite{BC1}. Chandrasekhar
\cite{SC2} showed that the massive Dirac's equation is also separable in the Kerr geometry
using the Newman-Penrose's null-tetrad formalism \cite{NP}. Subsequently his remarkable work
was further extended by Page and other people \cite{DKN} to the case of a Kerr-Newman black
hole \cite{KNm}.

Walker and Penrose \cite{WP} demonstrated that Carter's fourth constant, which is related to
the separability of the Hamilton-Jacobi equation in the Kerr background, is generated by an
irreducible second order symmetric St\"{a}ckel-Killing tensor that can be constructed out of
the Weyl tensor. Similarly, Carter and McLenaghan \cite{CM} found that the separability of
the Dirac equation in the Kerr geometry, is related to the fact that the skew-symmetric tensor
corresponding to the two-index Killing spinor admitted by the Kerr metric is a Killing-Yano
tensor of rank-two. Later then, a lot of efforts have been focused on showing that these
results can be extended to separability of higher-spin field equations in the Kerr spacetime
\cite{HSF}. For example, it has been shown \cite{KS} that Killing-Yano tensors and the Killing
spinor play a crucial role in separation of variables for the Maxwell's equation ($s = 1$),
Rarita-Schwinger's equation ($s = 3/2$), and the gravitational perturbation equation in the
Kerr geometry. These results have been extended to the more general classes of Petrov type-D
vacuum spacetime; see Ref. \cite{KMW} for a comprehensive review.

As was remarked by Chandrasekhar \cite{SC1}, the most striking feature of the Kerr metric is
the separability of all the standard wave equations in it. For some of these equations, their
separability has been understood as a consequence of the existence of certain tensor fields,
which have been found to be associated with a Killing spinor. In fact, it is just the existence
of various Killing objects (vectors, tensors and spinors) that ensures the separability of
arbitrary spin fields and the decoupling of Hamilton-Jacobi equation.

The separation of various equations can be understood in terms of different order differential
operators that characterize the separation constants appeared in the separable solutions. The
differential operators characterizing separation constants \cite{HSF} are also symmetry operators
of the various field equations in question. The essential property that allows the construction
of such operators is the existence of a Killing-Yano tensor in the Kerr spacetime. Physically,
Killing-Yano tensors and operators constructed from them have been associated with angular
momentum. In the case of the Kerr metric, which is of type D, the separation constant can be
characterized in terms of the Killing-Yano tensor admitted by the metric. In addition, it has
already been shown that many of the remarkable properties of the Kerr spacetime are consequences
of the existence of the Killing-Yano tensor, which means that all the symmetries responsible
for the separability of various equations are `derivable' from the Killing-Yano tensor \cite{CDC}.

In many aspects, a Killing-Yano tensor is more fundamental than a St\"{a}ckel-Killing tensor.
Namely, its square is always St\"{a}ckel-Killing tensor, but the opposite is not generally true.
The existence of a Killing-Yano tensor imposes additional conditions to those implied by the
existence of a two-index Killing spinor; whereas all the type-D vacuum metrics admit a two-index
Killing spinor, not all of them admit a Killing-Yano tensor. Nevertheless, in the case of
massless fields (particles), the separability of the corresponding equations is associated
with the existence of a two-index Killing spinor. In other words, the existence of a Killing
spinor implies that it is in relation to the separability of the massless field equation.

Higher dimensional generalizations of the famous Kerr black hole (with or without a cosmological
constant) and their properties have attracted a lot of interest \cite{HDBHs} in recent years,
in particular, with the discovery of the AdS/CFT correspondence. The vacuum solutions describing
the neutral rotating black holes in higher dimensions were constructed by Myers and Perry
\cite{MP} as the asymptotically flat generalizations of the Kerr metric. By introducing a
nonzero cosmological constant, Hawking \textit{et al}. \cite{HHT} obtained the asymptotically
nonflat generalizations in five dimensions with two independent angular momenta and in higher
dimensions with just one nonzero angular momentum parameter (see also \cite{Klemm}). Further
vacuum generalizations to all dimensions have been made recently in \cite{GLPPCLP}. Quite
recently, an exact charged generalization of the Kerr-Newman solution in five dimensions was
obtained in \cite{EMCS} within the framework of minimally gauged supergravity theory. Other
rotating charged black hole solutions in five-dimensional gauged and ungauged supergravity
were also obtained in \cite{CCLP,SUGBH,SEM,CYL,GodelBH}.

During the past years, a resurgence of interest \cite{FS,KF,KL,DKL,CGLP,VS,CLP,PD,KKAA,KKPV,FKK,HOY}
came about in the study of hidden symmetry and separability properties of the Klein-Gordon
scalar equation, the Hamilton-Jacobi equation, and stationary strings \cite{Sstring} of
higher-dimensional spacetimes \cite{CLHHOY} when Frolov and his collaborators (\cite{FS},
see \cite{VFR} for a review and references therein) showed that the five-dimensional
Myers-Perry metric possesses a lot of miraculous properties similar to the Kerr metric.
Specifically speaking, it has been shown that the Myers-Perry spacetime allows the separation
of variables in the geodesic Hamilton-Jacobi equation and the separability of the massless
Klein-Gordon scalar field equation \cite{FS}. These properties are also intimately connected
with the existence of a second order St\"{a}ckel-Killing tensor \cite{FS} admitted by the
five-dimensional Myers-Perry metric. It was further demonstrated that this rank-two
St\"{a}ckel-Killing tensor can be constructed from its ``square root'', a rank-three
Killing-Yano tensor \cite{KF}. Following a procedure put forward by Carter \cite{BC2},
Frolov \textit{et al}. \cite{KF} started from a potential one-form to generate a rank-two
conformal Killing-Yano tensor \cite{JL}, whose Hodge dual is just the expected Killing-Yano
tensor. Subsequently, these results have further been extended \cite{KF,KKPV,FKK,HOY} to
general higher-dimensional rotating black hole solutions with NUT charges \cite{GLPPCLP}.

On the other hand, there is relatively less work \cite{MS,OY,WuMP} to address the separability
of Dirac's equation and other higher-spin fields and its relation to the Killing-Yano tensor in
higher-dimensional rotating black holes. In a previous work \cite{WuMP}, we have investigated
the separability of a massive fermion field equation in the five-dimensional Myers-Perry
spacetime with two unequal angular momenta \cite{MP} and its relation to a Killing-Yano
tensor of rank-three. A first order symmetry operator commuting with the Dirac operator
has been constructed by using the rank-three Killing-Yano tensor whose square is just the
rank-two symmetric St\"{a}ckel-Killing tensor. In addition, we have obtained a second order
symmetry operator that commutes with the scalar Laplacian operator.

In this article, we shall extend that work to deal with the case of a nonzero cosmological
constant, namely, the five-dimensional Kerr-(anti-)de Sitter black holes with two independent
angular momenta \cite{HHT}. More specifically, we shall study the separation of variables
for a massive Klein-Gordon equation and a massive Dirac equation in this general rotating
five-dimensional black hole background spacetime. Symmetry operators that commutes respectively
with the scalar Laplacian operator and the standard Dirac operator are directly constructed
from the separated parts of these equations and are expressed in terms of the rank-two
St\"{a}ckel-Killing tensor and the rank-three Killing-Yano tensor.

This paper is outlined as follows. In Sec. \ref{mKAdS}, we present an elegant form for the
line element of the five-dimensional Kerr-(anti-)de Sitter black hole \cite{HHT} in the
Boyer-Lindquist coordinates. This new metric form allows us to explicitly construct the
local orthonormal coframe one-forms (pentad). we also rewrite the metric of the five-dimensional
Kerr-(anti-)de Sitter metric in a manner similar to the Plebanski solution \cite{PDle} in
four dimensions and give a brief review of the relevant symmetry properties of the spacetime.
In addition, we extend the first law of black hole thermodynamics to the case that the
cosmological constant is viewed as a thermodynamical variable. In Sec. \ref{KGSKT}, we
focus on the separation of variables for a massive Klein-Gordon equation in the background
and use the separated solutions to construct a concise expression for the St\"{a}ckel-Killing
tensor and a second order operator that commutes with the scalar Laplacian operator. Sec.
\ref{DEKAdS} is devoted to the separation of variables for a massive Dirac's equation in
the five-dimensional Kerr-(anti-)de Sitter black hole geometry. In this section, the
f\"{u}nfbein form of Dirac's field equation is adopted. Using Clifford algebra and the
spinor representation of SO(4,1), we construct the spinor connection one-forms. Here the
spinor connection is obtained by making use of the homomorphism between the SO(4,1) group
and its spinor representation which is derived from the Clifford algebra defined by the
anticommutation relations of the gamma matrices. Then the massive Dirac equation in a
five-dimensional Kerr-(anti-)de Sitter black hole is separated into purely radial and
purely angular equations. In Sec. \ref{OKYt}, we construct a first order symmetry operator
that commutes with the Dirac operator from the separated part of Dirac's equation. The
operator is explicitly expressed in terms of the rank-three Killing-Yano tensor (and its
covariant derivative) admitted by the five-dimensional Kerr-(anti-)de Sitter metric. The
last section \ref{CoRe} ends up with a brief summary of this paper and the related work
in progress. In the appendix, the affine spin-connection one-forms are calculated by the
first Cartan structure equation from the exterior differential of the pentad. The curvature
two-forms are also given in this pentad formalism.

\section{New form of the metric of a 5-dimensional Kerr-(anti-)de
Sitter black hole and its fundamental properties} \label{mKAdS}

The metric of a five-dimensional rotating black hole with two independent angular momenta
and a negative cosmological constant was given by Hawking \textit{et al}. \cite{HHT} in
1999. To construct a local orthonormal pentad with which the Dirac equation can be decoupled
into purely radial and purely angular parts, one hopes the metric for the five-dimensional
Kerr-(anti-)de Sitter black hole can be put into a local quasi-diagonal form. As did in Ref.
\cite{WuMP}, we find that the line element of the Kerr-(anti-)de Sitter metric \cite{HHT}
in five dimensions can be recast into a new form in the Boyer-Lindquist coordinates as follows:
\bea\hspace*{-1cm}
ds^2 &=& g_{\mu\nu}dx^{\mu}dx^{\nu} = \eta_{AB}e^A\otimes e^B \,  \nn \\
 &=& -\frac{\Delta_r}{\Sigma} X^2 +\frac{\Sigma}{\Delta_r}dr^2
  +\frac{\Sigma}{\Delta_{\theta}}d\theta^2
  +\frac{\Delta_{\theta}(a^2-b^2)^2\sin^2\theta\cos^2\theta}{p^2\Sigma} Y^2
  +\Big(\frac{ab}{rp} Z\Big)^2 \, , \quad
\label{KMP}
\eea
where
\numparts
\bea
X &=& dt -\frac{a\sin^2\theta}{\chi_a}d\phi -\frac{b\cos^2\theta}{\chi_b}d\psi \, , \\
Y &=& dt -\frac{(r^2+a^2)a}{(a^2-b^2)\chi_a}d\phi
 -\frac{(r^2+b^2)b}{(b^2-a^2)\chi_b}d\psi \, , \\
Z &=& dt -\frac{(r^2+a^2)\sin^2\theta}{a\chi_a}d\phi
 -\frac{(r^2+b^2)\cos^2\theta}{b\chi_b}d\psi \, ,
\eea
\endnumparts
and
\bea\hspace*{-2cm}
&& \Delta_r = (r^2+a^2)(r^2+b^2)\Big(\frac{1}{r^2} -\frac{\epsilon}{l^2}\Big) -2M \, , \qquad
 \chi_a = 1 +\epsilon\frac{a^2}{l^2} \, , \quad \chi_b = 1 +\epsilon\frac{b^2}{l^2} \, , \nn \\
\hspace*{-1cm}
&& \Delta_{\theta} = 1 +\epsilon\frac{p^2}{l^2} \, , \qquad \Sigma = r^2 +p^2 \, , \qquad
 p = \sqrt{a^2\cos^2\theta +b^2\sin^2\theta} \, . \nn
\eea
Here the symbol $\epsilon = 1$, $0$, and $-1$ corresponds to the Kerr-de Sitter, Kerr
(Myers-Perry), and Kerr-anti de Sitter cases, respectively. The parameters ($M, a, b, l$)
are related to the mass and two independent angular momenta of the black hole, and the
cosmological constant. Our conventions are as follows: Greek letters $\mu, \nu$ run over
five-dimensional spacetime coordinate indices $\{t, r, \theta, \phi, \psi\}$, while Latin
letters $A, B$ denote local orthonormal (Lorentz) frame indices $\{0, 1, 2, 3, 5\}$.
$\eta_{AB} = diag (-1, 1, 1, 1, 1)$ is the flat (Lorentz) metric tensor. Units are used
as $G = \hbar = c = 1$ throughout this paper.

Thermodynamics of $D = 5$ Kerr-anti de Sitter black holes had been studied \cite{BHT} in
details during the past years in the case of a fixed cosmological constant. In that case,
the integral Bekenstein-Smarr mass formulas can not be written as a closed form. On the
other hand, if the cosmological constant can be viewed as a thermodynamical variable
\cite{WuFL,CCK}, then both the differential and the integral mass formulae can be written
in a perfect form. Now we try to extend the black hole thermodynamics to the case of a
variable cosmological constant.

Note, however, that the above line element is written in a coordinate frame rotating at
infinity. To compute the physical mass and angular momenta, one has to change the metric
into a coordinate frame nonrotating at infinity by making the transformations: $\phi =
\widetilde{\phi} +\epsilon at/l^2$ and $\psi = \widetilde{\psi} +\epsilon bt/l^2$. After
doing so, we find that it only needs to make the following replacements in the line element
\numparts\bea\hspace*{-1cm}
X &=& \frac{1 +\epsilon p^2/l^2}{\chi_a\chi_b}dt
 -\frac{a\sin^2\theta}{\chi_a}d\widetilde{\phi}
 -\frac{b\cos^2\theta}{\chi_b}d\widetilde{\psi} \, , \\ \hspace*{-1cm}
Y &=& \frac{1 -\epsilon r^2/l^2}{\chi_a\chi_b}dt
 -\frac{(r^2+a^2)a}{(a^2-b^2)\chi_a}d\widetilde{\phi}
 -\frac{(r^2+b^2)b}{(b^2-a^2)\chi_b}d\widetilde{\psi} \, , \\ \hspace*{-1cm}
Z &=& \frac{(1 +\epsilon p^2/l^2)(1 -\epsilon r^2/l^2)}{\chi_a\chi_b}dt
 -\frac{(r^2+a^2)\sin^2\theta}{a\chi_a}d\widetilde{\phi}
 -\frac{(r^2+b^2)\cos^2\theta}{b\chi_b}d\widetilde{\psi} \, .
\eea\endnumparts

The outer event horizon is determined by the largest root of $\Delta_{r_+} = 0$. The Hawking
temperature $T = \kappa/(2\pi)$ and the Bekenstein-Hawking entropy $S = A/4$ with respect to
this horizon can be easily computed as
\be\hspace*{-1cm}
T = \frac{r_+^4\big[1 -\epsilon (2r_+^2 +a^2 +b^2)/l^2\big] -a^2b^2}{2\pi
 r_+(r_+^2 +a^2)(r_+^2 +b^2)} \, , \qquad
S = \pi^2\frac{(r_+^2 +a^2)(r_+^2+b^2)}{2\chi_a\chi_br_+} \, ,
\ee
while the angular velocities are measured as
\be
\Omega_a = \frac{a(1 -\epsilon r_+^2/l^2)}{r_+^2 +a^2} \, , \qquad
\Omega_b = \frac{b(1 -\epsilon r_+^2/l^2)}{r_+^2 +b^2} \, .
\ee

The physical mass and angular momenta are given by \cite{BHT}
\be
\mathcal{M} = \frac{\pi M}{2\chi_a\chi_b}\Big(\frac{1}{\chi_a}
 +\frac{1}{\chi_b} -\frac{1}{2}\Big) \, , \qquad
J_a = \frac{\pi Ma}{2\chi_a^2\chi_b} \, , \quad
J_b = \frac{\pi Mb}{2\chi_a\chi_b^2} \, ,
\ee
which obey the closed forms \cite{WuFL,CCK} for the first law of black hole thermodynamics
\bea
\frac{2}{3}\mathcal{M} &=& TS +\Omega_aJ_a +\Omega_bJ_b -\frac{1}{3}\Theta l \, , \\
d\mathcal{M} &=& TdS +\Omega_adJ_a +\Omega_bdJ_b -\Theta dl \, ,
\eea
where we have introduced the generalized force conjugate to the cosmological radius $l$ as
\bea
\Theta = \frac{\pi M}{2\chi_a\chi_b l}\Big(1 +\frac{1}{\chi_a} +\frac{1}{\chi_b}
 -\frac{3}{1 -\epsilon r_+^2/l^2}\Big) \, .
\eea
In the case without a cosmological constant, the above expressions reduce to the well-known
result given in \cite{MP,GMT} for the $D = 5$ Myers-Perry black holes.

In the practice of algebraic computations with the help of a computer program such as the
Maple-based GRTensor, one finds that it is much more efficient to use $p$ rather than $\theta$
itself as the appropriate angle coordinate. What is more, the radial part and the angular
part can be presented in a symmetric manner. In what follows, we shall adopt $p$ as the
convenient angle coordinate throughout this article. In doing so, the five-dimensional
Kerr-(anti-)de Sitter metric can be rewritten as
\be
ds^2 = -\frac{\Delta_r}{\Sigma} X^2 +\frac{\Sigma}{\Delta_r}dr^2
  +\frac{\Sigma}{\Delta_p}dp^2 +\frac{\Delta_p}{\Sigma} Y^2
  +\Big(\frac{ab}{rp} Z\Big)^2 \, ,
\label{mPDf}
\ee
where
\be
\Delta_p = -(p^2-a^2)(p^2-b^2)\Big(\frac{1}{p^2} +\frac{\epsilon}{l^2}\Big) \, ,
\ee
and
\numparts\bea
X &=& dt -\frac{(p^2-a^2)a}{(b^2-a^2)\chi_a}d\phi
 -\frac{(p^2-b^2)b}{(a^2-b^2)\chi_b}d\psi \, ,  \\
Y &=& dt +\frac{(r^2+a^2)a}{(b^2-a^2)\chi_a}d\phi
 +\frac{(r^2+b^2)b}{(a^2-b^2)\chi_b}d\psi\, , \\
Z &=& dt -\frac{(r^2+a^2)(p^2-a^2)}{(b^2-a^2)a\chi_a}d\phi
 -\frac{(r^2+b^2)(p^2-b^2)}{(a^2-b^2)b\chi_b}d\psi\, .
\eea\endnumparts

Performing the following coordinate transformations:
\be
t = \tau +(a^2+b^2)u +a^2b^2v \, , \quad \phi = a\chi_a(u +b^2v) \, , \quad
\psi = b\chi_b(u +a^2v) \, ,
\ee
we get
\bea
X = d\tau +p^2du \, , \quad Y = d\tau -r^2du \, , \quad
Z = d\tau +(p^2-r^2)du -r^2p^2dv \, ,
\eea
and find that the line element (\ref{mPDf}) of the $D = 5$ Kerr-(anti-)de Sitter metric is
very similar to the four-dimensional Plebanski solution \cite{PDle}.

The metric determinant for this spacetime is $\sqrt{-g} = rp\Sigma/[(a^2-b^2)\chi_a\chi_b]$,
and the contra-invariant metric tensor can be read accordingly from
\bea\hspace*{-1cm}
g^{\mu\nu}\p_{\mu}\p_{\nu} &=& \eta^{AB}\p_A\otimes\p_B \nn \\
&=& -\frac{(r^2+a^2)^2(r^2+b^2)^2}{r^4\Delta_r\Sigma}\Big(\p_t
 +\frac{a\chi_a}{r^2+a^2}\p_{\phi} +\frac{b\chi_b}{r^2+b^2}\p_{\psi}\Big)^2
 +\frac{\Delta_r}{\Sigma}\p_r^2 \nn \\
&& +\frac{\Delta_p}{\Sigma}\p_p^2 +\frac{(p^2-a^2)^2(p^2-b^2)^2}{p^4\Delta_p\Sigma}
\Big(\p_t -\frac{a\chi_a}{p^2-a^2}\p_{\phi} -\frac{b\chi_b}{p^2-b^2}\p_{\psi}\Big)^2 \nn \\
&&\quad +\frac{1}{r^2p^2}\big(ab\p_t +b\chi_a\p_{\phi} +a\chi_b\p_{\psi}\big)^2 \, .
\eea

The Kerr-(anti-)de Sitter metric (\ref{KMP}) possesses three Killing vectors ($\p_t$,
$\p_{\phi}$, and $\p_{\psi}$), In addition, it also admits a rank-two symmetric
St\"{a}ckel-Killing tensor, which can be written as the square of a rank-three Killing-Yano
tensor. It has been found that the existence of the St\"{a}ckel-Killing tensor ensures the
separation of variables in the geodesic Hamilton-Jacobi equation and the separability of
the massless Klein-Gordon scalar field equation. In this paper, we will show that the
separability of Dirac's equation in this spacetime background is also closely associated
with the existence of the rank-three Killing-Yano tensor.

The spacetime metric (\ref{KMP}) is of Petrov type-D \cite{PJDS,TypeD}. It possesses a pair
of real principal null vectors $\{\mathbf{l}, \mathbf{n}\}$, a pair of complex principal null
vectors $\{\mathbf{m}, \bar{\mathbf{m}}\}$, and one real, spatial-like unit vector $\mathbf{k}$.
They can be constructed to be of Kinnersley-type as follows:
\numparts\bea\hspace*{-1cm}
\mathbf{l}^{\mu}\p_{\mu} &=& \frac{(r^2+a^2)(r^2+b^2)}{r^2\Delta_r}\Big(\p_t
 +\frac{a\chi_a}{r^2+a^2}\p_{\phi} +\frac{b\chi_b}{r^2+b^2}\p_{\psi}\Big) +\p_r \, , \\ \hspace*{-1cm}
\mathbf{n}^{\mu}\p_{\mu} &=& \frac{(r^2+a^2)(r^2+b^2)}{2r^2\Sigma}\Big(\p_t
 +\frac{a\chi_a}{r^2+a^2}\p_{\phi} +\frac{b\chi_b}{r^2+b^2}\p_{\psi}\Big)
 -\frac{\Delta_r}{2\Sigma}\p_r \, , \\ \hspace*{-1cm}
\mathbf{m}^{\mu}\p_{\mu} &=& \frac{\sqrt{\Delta_p/2}}{r+ip}\bigg[\p_p
 +i\frac{(p^2-a^2)(p^2-b^2)}{p^2\Delta_p}\Big(\p_t
 -\frac{a\chi_a}{p^2-a^2}\p_{\phi} -\frac{b\chi_b}{p^2-b^2}\p_{\psi}\Big)\bigg] \, , \\ \hspace*{-1cm}
\bar{\mathbf{m}}^{\mu}\p_{\mu} &=& \frac{\sqrt{\Delta_p/2}}{r-ip}\bigg[\p_p
 -i\frac{(p^2-a^2)(p^2-b^2)}{p^2\Delta_p}\Big(\p_t
 -\frac{a\chi_a}{p^2-a^2}\p_{\phi} -\frac{b\chi_b}{p^2-b^2}\p_{\psi}\Big)\bigg] \, , \\ \hspace*{-1cm}
\mathbf{k}^{\mu}\p_{\mu} &=& \frac{1}{rp}\big(ab\p_t +b\chi_a\p_{\phi} +a\chi_b\p_{\psi}\big) \, .
\eea\endnumparts
These vectors are geodesic and satisfy the following orthogonal relations
\be
\mathbf{l}^{\mu}\mathbf{n}_{\mu} = -1 \, ,  \qquad
\mathbf{m}^{\mu}\bar{\mathbf{m}}_{\mu} = 1 \, ,
\qquad\quad \mathbf{k}^{\mu}\mathbf{k}_{\mu} = 1 \, ,
\label{Ognr}
\ee
and all others are zero. In terms of these vectors, the metric for the Kerr-(anti-)de Sitter
black hole (\ref{KMP}) can be put into a seminull pentad formalism ($2\bar{2}1$ formalism)
\cite{WuMP} as follows:
\be
ds^2 = -\mathbf{l}\otimes \mathbf{n} -\mathbf{n}\otimes \mathbf{l} +\mathbf{m}\otimes
 \bar{\mathbf{m}} +\bar{\mathbf{m}}\otimes \mathbf{m} +\mathbf{k}\otimes \mathbf{k} \, .
\ee

\section{St\"{a}ckel-Killing tensor and second order symmetry operator
from the separated solution of the massive Klein-Gordon equation} \label{KGSKT}

In this section, the massive Klein-Gordon scalar field equation is shown to be separable in
the five-dimensional Kerr-(anti-)de Sitter metric. From the separated solution of the radial
and angular parts, we construct a second order symmetry operator that commutes with the
scalar Laplacian operator. We then show that a second order, symmetric, St\"{a}ckel-Killing
tensor has a simple and elegant form in the local Lorentz pentad, which can be easily written
as the square of a rank-three Killing-Yano tensor.

To begin with, let us consider a massive Klein-Gordon scalar field equation
\be
\big(\Box -\mu_0^2\big)\Phi = \frac{1}{\sqrt{-g}}\p_{\mu}\big(\sqrt{-g}
 g^{\mu\nu}\p_{\nu}\Phi\big) -\mu_0^2\Phi = 0 \, ,
\ee
together with the ansatz $\Phi = R(r)S(p)e^{i(m\phi +k\psi -\omega t)}$. In the background
spacetime metric (\ref{KMP}), the massive scalar field equation reads
\bea\hspace*{-1cm}
&&\bigg\{ -\frac{(r^2+a^2)^2(r^2+b^2)^2}{r^4\Delta_r\Sigma}\Big(\p_t
 +\frac{a\chi_a}{r^2+a^2}\p_{\phi} +\frac{b\chi_b}{r^2+b^2}\p_{\psi}\Big)^2
 +\frac{1}{r\Sigma}\p_r\big(r\Delta_r\p_r\big) \nn \\ \hspace*{-1cm}
&&\qquad +\frac{1}{p\Sigma}\p_p\big(p\Delta_p\p_p\big)
 +\frac{(p^2-a^2)^2(p^2-b^2)^2}{p^4\Delta_p\Sigma}\Big(\p_t
 -\frac{a\chi_a}{p^2-a^2}\p_{\phi} -\frac{b\chi_b}{p^2-b^2}\p_{\psi}\Big)^2 \nn \\ \hspace*{-1cm}
&&\qquad\quad +\frac{1}{r^2p^2}\big(ab\p_t +b\chi_a\p_{\phi} +a\chi_b\p_{\psi}\big)^2
 -\mu_0^2\bigg \}\Phi = 0 \, .
\eea
Apparently, it can be separated into a radial part and an angular part,
\bea
&& \frac{1}{r}\p_r\big(r\Delta_r\p_rR\big)
 +\Big\{\frac{(r^2+a^2)^2(r^2+b^2)^2}{r^4\Delta_r}\Big(\omega
 -\frac{ma\chi_a}{r^2+a^2} -\frac{kb\chi_b}{r^2+b^2}\Big)^2 \nn \\
&&\qquad\qquad\qquad\quad  -\frac{1}{r^2}\big(ab\omega -mb\chi_a -ka\chi_b\big)^2
 -\mu_0^2r^2 -\lambda_0^2 \Big\}R(r) = 0 \, , \label{srs} \\ \hspace*{-1cm}
&& \frac{1}{p}\p_p\big(p\Delta_p\p_pS\big)
 -\Big\{\frac{(p^2-a^2)^2(p^2-b^2)^2}{p^4\Delta_p}\Big(\omega
 +\frac{ma\chi_a}{p^2-a^2} +\frac{kb\chi_b}{p^2-b^2}\Big)^2 \nn \\
&&\qquad\qquad\qquad\quad  +\frac{1}{p^2}\big(ab\omega -mb\chi_a -ka\chi_b\big)^2
 +\mu_0^2p^2 -\lambda_0^2\Big\}S(p) = 0 \, . \label{sra}
\eea
Both of them can be transformed into the general form of Heun equation \cite{KL,PD,Heun}.

Now from the separated parts (\ref{srs}) and (\ref{sra}), one can construct a new dual equation
as follows:
\bea\hspace*{-1cm}
&&\bigg\{ -p^2\frac{(r^2+a^2)^2(r^2+b^2)^2}{r^4\Delta_r\Sigma}\Big(\p_t
 +\frac{a\chi_a}{r^2+a^2}\p_{\phi} +\frac{b\chi_b}{r^2+b^2}\p_{\psi}\Big)^2
 +\frac{p^2}{r\Sigma}\p_r\big(r\Delta_r\p_r\big) \nn \\ \hspace*{-1cm}
&&\qquad -\frac{r^2}{p\Sigma}\p_p\big(p\Delta_p\p_p\big)
 -r^2\frac{(p^2-a^2)^2(p^2-b^2)^2}{p^4\Delta_p\Sigma}\Big(\p_t
 -\frac{a\chi_a}{p^2-a^2}\p_{\phi} -\frac{b\chi_b}{p^2-b^2}\p_{\psi}\Big)^2 \nn \\ \hspace*{-1cm}
&&\qquad\quad +\frac{p^2-r^2}{r^2p^2}\big(ab\p_t +b\chi_a\p_{\phi}
b+a\chi_b\p_{\psi}\big)^2 -\lambda_0^2\bigg \}\Phi = 0 \, ,
\eea
from which one can extract a second order symmetric tensor --- the so-called St\"{a}ckel-Killing
tensor
\bea\hspace*{-1cm}
K^{\mu\nu}\p_{\mu}\p_{\nu} &=& -p^2\frac{(r^2+a^2)^2(r^2+b^2)^2}{r^4\Delta_r\Sigma}
 \Big(\p_t +\frac{a\chi_a}{r^2+a^2}\p_{\phi} +\frac{b\chi_b}{r^2+b^2}\p_{\psi}\Big)^2
 +p^2\frac{\Delta_r}{\Sigma}\p_r^2 \nn \\
&& -r^2\frac{\Delta_p}{\Sigma}\p_p^2
 -r^2\frac{(p^2-a^2)^2(p^2-b^2)^2}{p^4\Delta_p\Sigma}\Big(\p_t
 -\frac{a\chi_a}{p^2-a^2}\p_{\phi} -\frac{b\chi_b}{p^2-b^2}\p_{\psi}\Big)^2 \nn \\
&&\quad +\frac{p^2-r^2}{r^2p^2}\big(ab\p_t +b\chi_a\p_{\phi} +a\chi_b\p_{\psi}\big)^2 \, .
\label{5dKT}
\eea
This symmetric tensor $K_{\mu\nu} = K_{\nu\mu}$ obeys the Killing equation \cite{WP}
\be
K_{\mu\nu;\rho} +K_{\nu\rho;\mu} +K_{\rho\mu;\nu} = 0 \, ,
\label{Kte}
\ee
and is equivalent to those found in \cite{FS,KF}, up to an additive constant. In the local
Lorentz coframe given in Eq. (\ref{pentad}), it has a simple, diagonal form
\be
K_{AB} = \mbox{diag} (-p^2, p^2, -r^2, -r^2, p^2-r^2) \, .
\ee

Using the St\"{a}ckel-Killing tensor, the above dual equation can be written in a
coordinate-independent form
\be
\big(\mathbb{K} -\lambda_0^2\big)\Phi = \frac{1}{\sqrt{-g}}\p_{\mu}\big(\sqrt{-g}
 K^{\mu\nu}\p_{\nu}\Phi\big) -\lambda_0^2\Phi = 0 \, .
\ee
Clearly, the symmetry operator $\mathbb{K}$ is expressed in terms of the St\"{a}ckel-Killing
tensor and commutes with the scalar Laplacian operator $\Box$. Expanding the commutator
$[\mathbb{K}, \Box] = 0$ yields the Killing equation (\ref{Kte}) and the integrability
condition for the St\"{a}ckel-Killing tensor. These two operators have a classical analogue.
In classical mechanics, the scalar Laplacian operator $\Box$ corresponds to the Hamiltonian
$g_{\mu\nu}\dot{x}^{\mu}\dot{x}^{\nu}$, while the operator $\mathbb{K}$ to the Carter's
constant $K_{\mu\nu}\dot{x}^{\mu}\dot{x}^{\nu}$. They are two integrals of motion in addition
to three constants induced from the Killing vector fields $\p_t$, $\p_{\phi}$, and $\p_{\psi}$.

\section{Separability of the massive Dirac field equation in a
5-dimensional Kerr-(anti-)de Sitter black hole} \label{DEKAdS}

In Ref. \cite{WuMP}, the Dirac equation for spin-$1/2$ fermions in the general Myers-Perry
black hole geometry has been decoupled into purely radial and purely angular parts by using
the orthonormal f\"{u}nfbein (pentad) formalism in the five-dimensional relativity. In this
section, we shall extend that work to the case of a nonzero cosmological constant. That is,
we will still work out the Dirac equation within a local orthonormal pentad formalism and
show that the Dirac equation is separable by variables in the $D = 5$ Kerr-(anti-)de Sitter
black hole geometry.

\subsection{F\"{u}nfbein formalism of Dirac field equation}

In curved background spacetime, the Dirac equation for the spinor field is
\be
\big(\mathbb{H}_D +\mu_e\big)\Psi =
 \big[\gamma^Ae_A^{~\mu}(\p_{\mu} +\Gamma_{\mu}) +\mu_e\big]\Psi = 0 \, ,
\label{DE}
\ee
where $\psi$ is a four-component Dirac spinor, $\mu_e$ is the mass of the electron,
$e_A^{~\mu}$ is the f\"{u}nfbein (pentad), its inverse $e_{~\mu}^A$ is defined by
$g_{\mu\nu} = \eta_{AB}e_{~\mu}^Ae_{~\nu}^B$, $\Gamma_{\mu}$ is the spinor connection,
and $\gamma^A$'s are the five-dimensional gamma matrices obeying the anticommutation
relations (Clifford algebra)
\be
\big\{\gamma^A, \gamma^B\big\} \equiv \gamma^A\gamma^B +\gamma^B\gamma^A = 2\eta^{AB} \, .
\label{Clifford}
\ee
For our purpose, we choose the following explicit representations for the gamma matrices
\cite{WuMP}
\bea
&& \gamma^0 = i\sigma^1\otimes I \, , \qquad\quad \gamma^1 = -i\sigma^2\otimes \sigma^3 \, ,
\qquad\quad \gamma^2 = -i\sigma^2\otimes \sigma^1 \, , \nn \\
&& \gamma^3 = -i\sigma^2\otimes \sigma^2 \, , \qquad\qquad\quad
\gamma^5 = \sigma^3\otimes I = -i\gamma^0\gamma^1\gamma^2\gamma^3 \, ,
\label{GMr}
\eea
where $\sigma^i$'s are the Pauli matrices, and $I$ is a $2 \times 2$ identity matrix,
respectively.

In order to derive the spinor connection one-forms $\Gamma = \Gamma_{\mu}dx^{\mu}\equiv
\Gamma_Ae^A$, we first compute the spin-connection one-forms $\omega_{AB} = \omega_{AB\mu}
dx^{\mu}\equiv f_{ABC}e^C$ in the orthonormal pentad coframe, i.e., the one-forms $e^A =
e^A_{~\mu}dx^{\mu}$ satisfying the torsion-free condition --- Cartan's first structure
equation and the skew-symmetric condition
\be
de^A +\omega^A_{~B}\wedge e^B = 0 \, , \qquad\qquad
\omega_{AB} = \eta_{AC}\omega_{~B}^C = -\omega_{BA} \, .
\label{CFE}
\ee
To obtain the spinor connection one-forms $\Gamma$ from $\omega_{AB}$, one can utilize the
homomorphism between the SO(4,1) group and its spinor representation derivable from the
Clifford algebra (\ref{Clifford}). The SO(4,1) Lie algebra is defined by the ten antisymmetric
generators $\Sigma^{AB} = [\gamma^A, \gamma^B]/(2i)$ which gives the spinor representation,
and the spinor connection $\Gamma$ can be regarded as a SO(4,1) Lie-algebra-valued one-form.
Using the isomorphism between the SO(4,1) Lie algebra and its spinor representation, i.e.,
$\Gamma_{\mu} = (i/4)\Sigma^{AB}\omega_{AB\mu} = (1/4)\gamma^A\gamma^B\omega_{AB\mu}$, one
can immediately construct the spinor connection one-forms
\be
\Gamma = \frac{1}{8}\big[\gamma^A, \gamma^B\big]\omega_{AB}
= \frac{1}{4}\gamma^A\gamma^B\omega_{AB} = \frac{1}{4}\gamma^A\gamma^Bf_{ABC}e^C \, .
\ee
Now in terms of the local differential operator $\p_A = e_A^{~\mu}\p_{\mu}$, the Dirac equation
(\ref{DE}) can be rewritten in the local Lorentz frame as
\be
\big(\mathbb{H}_D +\mu_e\big)\Psi = \big[\gamma^A(\p_A +\Gamma_A) +\mu_e\big]\Psi = 0 \, ,
\label{DELF}
\ee
where $\Gamma_A = e_A^{~\mu}\Gamma_{\mu} = (1/4)\gamma^B\gamma^Cf_{BCA}$ is the component
of the spinor connection in the local Lorentz frame. Note that the five-dimensional Clifford
algebra has two different but reducible representations which can differ by the multiplier
of a $\gamma^5$ matrix. It is usually assumed that fermion fields are in a reducible
representation of the Clifford algebra. In other words, one can work with the Dirac
equation in a four-component spinor formalism just like in the four-dimensional case,
and takes the $\gamma^5$ matrix as the fifth component of Clifford vectors.

\subsection{Computation of covariant spinor differential operator}

In the local Lorentz form of Dirac's equation, we need to find the local partial differential
operator $\p_A = e_A^{~\mu}\p_{\mu}$ and the spinor connection $\Gamma_A = e_A^{~\mu}
\Gamma_{\mu}$ subject to the Kerr-(anti-)de Sitter metric (\ref{KMP}). The orthonormal
basis one-vectors $\p_A $ dual to the pentad $e^{A}$ given in the Appendix Eq. (\ref{pentad})
are
\bea
&& \p_0 = \frac{(r^2+a^2)(r^2+b^2)}{r^2\sqrt{\Delta_r\Sigma}}\Big(\p_t
 +\frac{a\chi_a}{r^2+a^2}\p_{\phi} +\frac{b\chi_b}{r^2+b^2}\p_{\psi}\Big) \, , \nn \\
&& \p_1 = \sqrt{\frac{\Delta_r}{\Sigma}}\p_r \, , \qquad\qquad
 \p_2 = \sqrt{\frac{\Delta_p}{\Sigma}}\p_p \, , \nn \\
&& \p_3 = \frac{(p^2-a^2)(p^2-b^2)}{p^2\sqrt{\Delta_p\Sigma}}\Big(\p_t
 -\frac{a\chi_a}{p^2-a^2}\p_{\phi} -\frac{b\chi_b}{p^2-b^2}\p_{\psi}\Big) \, , \nn \\
&& \p_5 = \frac{1}{rp}\big(ab\p_t +b\chi_a\p_{\phi} +a\chi_b\p_{\psi}\big) \, .
\eea

Taking use of the local Lorentz frame component $\Gamma_A$ given in Eq. (\ref{Lsc}) and the
properties of gamma matrices together with the relation $\gamma^5 = -i\gamma^0\gamma^1\gamma^2
\gamma^3$, we obtain
\bea
\gamma^A\Gamma_A &=& \frac{1}{4}\gamma^A\gamma^B\gamma^Cf_{BCA} \nn \\
&=& \gamma^1\sqrt{\frac{\Delta_r}{\Sigma}}\Big(\frac{\Delta_r^{\prime}}{4\Delta_r}
 +\frac{1}{2r} +\frac{r}{2\Sigma} \Big) +\gamma^2\sqrt{\frac{\Delta_p}{\Sigma}}
 \Big(\frac{\Delta_p^{\prime}}{4\Delta_p} +\frac{1}{2p} +\frac{p}{2\Sigma}\Big) \nn \\
&& +\frac{r}{2\Sigma}\sqrt{\frac{\Delta_p}{\Sigma}}\gamma^0\gamma^1\gamma^3
 -\frac{ab}{2r^2p}\gamma^0\gamma^1\gamma^5
 +\frac{p}{2\Sigma}\sqrt{\frac{\Delta_r}{\Sigma}}\gamma^0\gamma^2\gamma^3
 +\frac{ab}{2rp^2}\gamma^2\gamma^3\gamma^5 \nn \\
&=& \gamma^1\sqrt{\frac{\Delta_r}{\Sigma}}\Big(\frac{\Delta_r^{\prime}}{4\Delta_r}
 +\frac{1}{2r} +\frac{r -ip\gamma^5}{2\Sigma} \Big) +\gamma^2\sqrt{\frac{\Delta_p}{\Sigma}}
 \Big(\frac{\Delta_p^{\prime}}{4\Delta_p} +\frac{1}{2p} \nn \\
&& +\frac{p +ir\gamma^5}{2\Sigma}\Big)
  +\frac{iab}{2r^2p^2}\gamma^0\gamma^1\big(r +ip\gamma^5\big) \, ,
\eea
where a prime denotes the partial differential with respect to the coordinates $r$ and $p$.

Combining the above expression with the spinor differential operator
\bea\hspace*{-1cm}
&& \gamma^A\p_A = \gamma^0\frac{(r^2+a^2)(r^2+b^2)}{r^2\sqrt{\Delta_r\Sigma}}\Big(\p_t
 +\frac{a\chi_a}{r^2+a^2}\p_{\phi} +\frac{b\chi_b}{r^2+b^2}\p_{\psi}\Big)
 +\gamma^1\sqrt{\frac{\Delta_r}{\Sigma}}\p_r \nn \\ \hspace*{-1cm}
&&\qquad\quad +\gamma^2\sqrt{\frac{\Delta_p}{\Sigma}}\p_p
 +\gamma^3 \frac{(p^2-a^2)(p^2-b^2)}{p^2\sqrt{\Delta_p\Sigma}}\Big(\p_t
 -\frac{a\chi_a}{p^2-a^2}\p_{\phi} -\frac{b\chi_b}{p^2-b^2}\p_{\psi}\Big) \nn \\ \hspace*{-1cm}
&&\qquad\qquad +\gamma^5\frac{1}{rp}\big(ab\p_t +b\chi_a\p_{\phi} +a\chi_b\p_{\psi}\big) \, ,
\eea
we find that the covariant Dirac differential operator in the local Lorentz frame is
\bea\hspace*{-1cm}
\mathbb{H}_D &=& \gamma^A(\p_A +\Gamma_A)
 = \gamma^0\frac{(r^2+a^2)(r^2+b^2)}{r^2\sqrt{\Delta_r\Sigma}}\Big(\p_t
 +\frac{a\chi_a}{r^2+a^2}\p_{\phi} +\frac{b\chi_b}{r^2+b^2}\p_{\psi}\Big) \nn \\ \hspace*{-1cm}
&&\quad +\gamma^1\sqrt{\frac{\Delta_r}{\Sigma}}\Big(\p_r +\frac{\Delta_r^{\prime}}{4\Delta_r}
 +\frac{1}{2r} +\frac{r -ip\gamma^5}{2\Sigma}\Big) +\gamma^2\sqrt{\frac{\Delta_p}{\Sigma}}
 \Big(\p_p +\frac{\Delta_p^{\prime}}{4\Delta_p} +\frac{1}{2p} \nn \\ \hspace*{-1cm}
&&\quad +\frac{p +ir\gamma^5}{2\Sigma}\Big)
 +\gamma^3\frac{(p^2-a^2)(p^2-b^2)}{p^2\sqrt{\Delta_p\Sigma}}\Big(\p_t
 -\frac{a\chi_a}{p^2-a^2}\p_{\phi} -\frac{b\chi_b}{p^2-b^2}\p_{\psi}\Big) \nn \\ \hspace*{-1cm}
&&\qquad +\gamma^5\frac{1}{rp}\big(ab\p_t +b\chi_a\p_{\phi} +a\chi_b\p_{\psi}\big)
  +\frac{iab}{2r^2p^2}\gamma^0\gamma^1\big(r +ip\gamma^5\big) \, .
\eea

\subsection{Separation of variables in Dirac equation}

With the above preparation in hand, we are in a position to decouple the Dirac equation.
Substituting the above spinor differential operator into Eq. (\ref{DELF}), the Dirac equation
in the five-dimensional Kerr-(anti-)de Sitter metric reads
\bea\hspace*{-1cm}
&& \bigg\{ \gamma^0\frac{(r^2+a^2)(r^2+b^2)}{r^2\sqrt{\Delta_r\Sigma}}\Big(\p_t
 +\frac{a\chi_a}{r^2+a^2}\p_{\phi} +\frac{b\chi_b}{r^2+b^2}\p_{\psi}\Big)
 +\gamma^1\sqrt{\frac{\Delta_r}{\Sigma}}\Big(\p_r
 +\frac{\Delta_r^{\prime}}{4\Delta_r} \nn \\ \hspace*{-1cm}
&&\qquad +\frac{1}{2r} +\frac{r -ip \gamma^5}{2\Sigma}\Big)
 +\gamma^2\sqrt{\frac{\Delta_p}{\Sigma}}\Big[\p_p +\frac{\Delta_p^{\prime}}{4\Delta_p}
 +\frac{1}{2p} +\frac{i\gamma^5}{2\Sigma}\big(r -ip\gamma^5\big)\Big] \nn \\ \hspace*{-1cm}
&&\qquad +\gamma^3\frac{(p^2-a^2)(p^2-b^2)}{p^2\sqrt{\Delta_p\Sigma}}
 \Big(\p_t -\frac{a\chi_a}{p^2-a^2}\p_{\phi}
 -\frac{b\chi_b}{p^2-b^2}\p_{\psi}\Big) \nn \\ \hspace*{-1cm}
&&\qquad  +\gamma^5\frac{1}{rp}\big(ab\p_t +b\chi_a\p_{\phi} +a\chi_b\p_{\psi}\big)
 +\frac{iab}{2r^2p^2}\gamma^0\gamma^1 \big(r +ip\gamma^5\big) +\mu_e \bigg\}\Psi = 0 \, .
\eea
Multiplying $(r -ip\gamma^5)\sqrt{r +ip\gamma^5} = \sqrt{\Sigma(r -ip\gamma^5)}$ by the left
to the above equation, and after some lengthy algebra manipulations we arrive at
\bea\hspace*{-1.5cm}
&& \bigg\{ \gamma^0\frac{(r^2+a^2)(r^2+b^2)}{r^2\sqrt{\Delta_r}}\Big(\p_t
 +\frac{a\chi_a}{r^2+a^2}\p_{\phi} +\frac{b\chi_b}{r^2+b^2}\p_{\psi}\Big)
 +\gamma^1\sqrt{\Delta_r}\Big(\p_r +\frac{\Delta_r^{\prime}}{4\Delta_r}
 +\frac{1}{2r}\Big) \nn \\ \hspace*{-1.5cm}
&&\quad +\gamma^2\sqrt{\Delta_p}\Big(\p_p +\frac{\Delta_p^{\prime}}{4\Delta_p}
 +\frac{1}{2p}\Big) +\gamma^3\frac{(p^2-a^2)(p^2-b^2)}{p^2\sqrt{\Delta_p}}\Big(\p_t
 -\frac{a\chi_a}{p^2-a^2}\p_{\phi} -\frac{b\chi_b}{p^2-b^2}\p_{\psi}\Big) \nn \\ \hspace*{-1.5cm}
&&\quad +\Big(\frac{\gamma^5}{p} -\frac{i}{r}\Big)\big(ab\p_t +b\chi_a\p_{\phi}
 +a\chi_b\p_{\psi}\big) +\Big(\frac{ab}{2p^2}
 +\frac{ab}{2r^2}\Big)i\gamma^0\gamma^1 \nn \\ \hspace*{-1.5cm}
&&\qquad  +\mu_e\big(r -ip\gamma^5\big) \bigg\}\big(\sqrt{r +ip\gamma^5}\Psi\big) = 0 \, .
\label{presde}
\eea

Now we assume that the spin-$1/2$ fermion fields are in a reducible representation of the
Clifford algebra and can be taken as a four-component Dirac spinor. Applying the explicit
representation (\ref{GMr}) for the gamma matrices and adopting the following ansatz for the
separation of variables
\be
\sqrt{r +ip\gamma^5}\Psi = e^{i(m\phi+k\psi-\omega t)}\left(\ba{cl}
&\hspace*{-10pt} R_2(r)S_1(p) \\
&\hspace*{-10pt} R_1(r)S_2(p) \\
&\hspace*{-10pt} R_1(r)S_1(p) \\
&\hspace*{-10pt} R_2(r)S_2(p)
\ea\right) \, ,
\ee
we find that the Dirac equation in the five-dimensional Kerr-(anti-)de Sitter metric can be
decoupled into the purely radial parts and the purely angular parts
\bea
&& \sqrt{\Delta_r}\cD_r^-R_1 = \Big[\lambda +i\mu_er -\frac{ab}{2r^2}
 -\frac{i}{r}\big(ab\omega -mb\chi_a -ka\chi_b\Big)\big]R_2 \, , \label{sdea} \\
&& \sqrt{\Delta_r}\cD_r^+R_2 = \Big[\lambda -i\mu_er -\frac{ab}{2r^2}
+\frac{i}{r}\big(ab\omega -mb\chi_a -ka\chi_b\big)\Big]R_1 \, , \label{sdeb} \\
&& \sqrt{\Delta_p}\cL_p^+S_1 = \Big[\quad\lambda +\mu_ep +\frac{ab}{2p^2}
 +\frac{1}{p}\big(ab\omega -mb\chi_a -ka\chi_b\big)\Big]S_2 \, , \label{sdec} \\
&& \sqrt{\Delta_p}\cL_p^-S_2 = \Big[-\lambda +\mu_ep -\frac{ab}{2p^2}
 +\frac{1}{p}\big(ab\omega -mb\chi_a -ka\chi_b\big)\Big]S_1 \label{sded} \, ,
\eea
in which we have denoted
\bea
&& \cD_r^{\pm} = \p_r +\frac{\Delta_r^{\prime}}{4\Delta_r} +\frac{1}{2r} \pm
 i\frac{(r^2+a^2)(r^2+b^2)}{r^2\Delta_r}\Big(\omega
 -\frac{ma\chi_a}{r^2+a^2} -\frac{kb\chi_b}{r^2+b^2}\Big) \, , \nn \\
&& \cL_p^{\pm} = \p_p +\frac{\Delta_p^{\prime}}{4\Delta_p} +\frac{1}{2p} \pm
  \frac{(p^2-a^2)(p^2-b^2)}{p^2\Delta_p}\Big(\omega +\frac{ma\chi_a}{p^2-a^2}
  +\frac{kb\chi_b}{p^2-b^2}\Big) \, . \nn
\eea

The separated radial and angular equations (\ref{sdea}-\ref{sded}) can be reduced into a
master equation containing only one component. For the radial part, they are explicitly
given by
\bea\hspace*{-1.5cm}
&& \frac{1}{r}\sqrt{\Delta_r}D_r\Big(r\sqrt{\Delta_r}D_rR_1\Big)
 +\bigg\{\frac{(r^2+a^2)^2(r^2+b^2)^2}{r^4\Delta_r}\Big(\omega
 -\frac{ma\chi_a}{r^2+a^2} -\frac{kb\chi_b}{r^2+b^2}\Big)^2 \nn \\ \hspace*{-1.5cm}
&&\quad -\frac{(ab\omega -mb\chi_a -ka\chi_b)^2}{r^2} +2\mu_e(ab\omega
 -mb\chi_a -ka\chi_b) -\mu_e^2r^2 -\lambda^2 +\lambda\frac{ab}{r^2} \nn \\ \hspace*{-1.5cm}
&&\quad -\frac{a^2b^2}{4r^4} -\frac{\lambda +2i\mu_er +ab/(2r^2)}{\lambda r
 +i\mu_er^2 -ab/(2r) -i\big(ab\omega -mb\chi_a -ka\chi_b\big)}\Delta_rD_r
 +\Big[\frac{2i}{r} +i\frac{\Delta_r^{\prime}}{2\Delta_r} \nn \\ \hspace*{-1.5cm}
&&\quad -\frac{\mu_er -iab/r^2 +(ab\omega -mb\chi_a -ka\chi_b)/r}{\lambda r
 +i\mu_er^2 -ab/(2r) -i(ab\omega -mb\chi_a -ka\chi_b)}\Big]
 \frac{(r^2+a^2)(r^2+b^2)}{r^2}\Big(\omega -\frac{ma\chi_a}{r^2+a^2} \nn \\ \hspace*{-1.5cm}
&&\qquad\quad -\frac{kb\chi_b}{r^2+b^2}\Big) -\frac{2i}{r}\Big[(2r^2+a^2+b^2)\omega
 -ma\chi_a -kb\chi_b\Big] \bigg\}R_1 = 0 \, ,
\eea
and
\bea\hspace*{-1.5cm}
&& \frac{1}{r}\sqrt{\Delta_r}D_r\Big(r\sqrt{\Delta_r}D_rR_2\Big)
 +\bigg\{\frac{(r^2+a^2)^2(r^2+b^2)^2}{r^4\Delta_r}\Big(\omega
 -\frac{ma\chi_a}{r^2+a^2} -\frac{kb\chi_b}{r^2+b^2}\Big)^2 \nn \\ \hspace*{-1.5cm}
&&\quad -\frac{(ab\omega -mb\chi_a -ka\chi_b)^2}{r^2} +2\mu_e(ab\omega
 -mb\chi_a -ka\chi_b) -\mu_e^2r^2 -\lambda^2 +\lambda\frac{ab}{r^2} \nn \\ \hspace*{-1.5cm}
&&\quad -\frac{a^2b^2}{4r^4} -\frac{\lambda -2i\mu_er +ab/(2r^2)}{\lambda r
 -i\mu_er^2 -ab/(2r) +i(ab\omega -mb\chi_a -ka\chi_b)}\Delta_rD_r
 -\Big[\frac{2i}{r} +i\frac{\Delta_r^{\prime}}{2\Delta_r} \nn \\ \hspace*{-1.5cm}
&&\quad +\frac{\mu_er +iab/r^2 +(ab\omega -mb\chi_a -ka\chi_b)/r}{\lambda r
-i\mu_er^2 -ab/(2r) +i(ab\omega -mb\chi_a -ka\chi_b)}\Big]
 \frac{(r^2+a^2)(r^2+b^2)}{r^2}\Big(\omega -\frac{ma\chi_a}{r^2+a^2} \nn \\ \hspace*{-1.5cm}
&&\qquad\quad -\frac{kb\chi_b}{r^2+b^2}\Big) +\frac{2i}{r}\Big[(2r^2+a^2+b^2)\omega
 -ma\chi_a -kb\chi_b\Big] \bigg\}R_2 = 0 \, ,
\eea
where
$$D_r = \p_r +\frac{\Delta_r^{\prime}}{4\Delta_r} +\frac{1}{2r} \, . $$

From the above decoupled master equations, it is easy to see that they are more complicated
than the four-dimensional case \cite{KBC,STUBS}. As for the exact solution to these equations,
one hopes that they can be transformed into the general form of Heun equation \cite{Heun},
like the four-dimensional case \cite{STUBS}. The case occurs similarly for the angular parts.
Moreover, the angular part can be transformed into the radial part if we replace $p$ by $ir$
in the vacuum case where $M = 0$.

\section{Construction of a first order symmetry operator
in terms of the Killing-Yano tensor} \label{OKYt}

In the last section, we have explicitly shown that Dirac's equation is separable in the $D = 5$
Kerr-(anti-)de Sitter black hole spacetime. In this section, we will demonstrate that this
separability is closely related to the existence of a rank-three Killing-Yano tensor admitted
by the Kerr-(anti-)de Sitter metric. To this end, we will construct a first order symmetry
operator that commutes with the Dirac operator by using the Killing-Yano tensor of rank-three.
This symmetry operator is directly constructed from the separated solutions of the Dirac's
equation.

\subsection{Killing-Yano potential and
(conformal) Killing-Yano tensor}

Before constructing a first order symmetry operator that commutes with the Dirac operator, we
first give a brief review on the recent work \cite{KF,KKPV,FKK} about the construction of the
St\"{a}ckel-Killing tensor from the (conformal) Killing-Yano tensor.

Penrose and Floyd \cite{PF} discovered that the St\"{a}ckel-Killing tensor for the 4-dimensional
Kerr metric can be written in the form $K_{\mu\nu} = f_{\mu\rho}f^{~\rho}_{\nu}$, where the
skew-symmetric tensor $f_{\mu\nu} = -f_{\nu\mu}$ is the Killing-Yano tensor \cite{KY,TK,DR}
obeying the equation $f_{\mu\nu;\rho} +f_{\mu\rho;\nu} = 0$. Using this object, Carter and
McLenaghan \cite{CM} constructed a first order symmetry operator that commutes with the massive
Dirac operator. In the case of a $D = 4$ Kerr black hole, the Killing-Yano tensor $f$ is of
rank-two, its Hodge dual $k = -{^*}f$ is a rank-two, antisymmetric, conformal Killing-Yano
tensor \cite{JL} obeying the equation
\be
k_{\alpha\beta;\gamma} +k_{\alpha\gamma;\beta} = \frac{1}{D-1}\big(g_{\alpha\beta}
 k^{\mu}_{~\gamma;\mu} +g_{\gamma\alpha}k^{\mu}_{~\beta;\mu} -2g_{\beta\gamma}
 k^{\mu}_{~\alpha;\mu}\big) \, .
\ee
This equation is equivalent to the Penrose's equation \cite{PenE}
\be
\mathcal{P}_{\alpha\beta\gamma} = k_{\alpha\beta;\gamma}
 +\frac{1}{D-1}\big(g_{\beta\gamma}k^{\mu}_{~\alpha;\mu}
 -g_{\gamma\alpha}k^{\mu}_{~\beta;\mu}\big) = 0 \, .
\ee

A conformal Killing-Yano tensor $k$ is dual to the Killing-Yano tensor if and only if it is
closed $dk = 0$. This fact implies that there exists a potential one-form $\hat{b}$ so that
$k = d\hat{b}$. Carter \cite{BC2} first found this potential to generate the Killing-Yano
tensor for the Kerr-Newman black hole.

Recently, these results have further been extended \cite{KF,KKPV,FKK,HOY} to higher-dimensional
rotating black hole solutions. In the case of $D = 5$ dimensions, it was demonstrated \cite{KF}
that the rank-two St\"{a}ckel-Killing tensor can be constructed from its ``square root'', a
rank-three, totally antisymmetric Killing-Yano tensor. Following Carter's procedure \cite{BC2},
Frolov \textit{et al}. \cite{KF} found a potential one-form to generate a rank-two conformal
Killing-Yano tensor \cite{JL}, whose Hodge dual $f = {^*}k$ is the expected rank-three
Killing-Yano tensor.

Now we focus on the case of the five-dimensional Kerr-(anti-)de Sitter metric. It is easy to
check that the following object constructed from the rank-three Killing-Yano tensor
\be
K_{\mu\nu} = -\frac{1}{2}f_{\mu\alpha\beta}f_{\nu}^{~\alpha\beta} \, ,
\ee
is just the rank-two, St\"{a}ckel-Killing tensor given in Eq. (\ref{5dKT}). The rank-three
Killing-Yano tensor $f$ obeying the equation
\be
f_{\alpha\beta\mu;\nu} +f_{\alpha\beta\nu;\mu} = 0 \, ,
\label{KYe}
\ee
can be taken as the Hodge dual $f = {^*}k$ of the 2-form $k = d\hat{b}$ via the following
definition:
\be
f_{\alpha\beta\gamma} = ({^*}k)_{\alpha\beta\gamma} = \frac{1}{2}
 \sqrt{-g}\varepsilon_{\alpha\beta\gamma\mu\nu}k^{\mu\nu}\, .
\ee

The Killing-Yano potential found for the five-dimensional Kerr-(anti-)de Sitter metric is
\cite{KF,WuMP}
\be
2\hat{b} = \big(p^2-r^2\big)dt +\frac{(r^2+a^2)(p^2-a^2)a}{(b^2-a^2)\chi_a}d\phi
 +\frac{(r^2+b^2)(p^2-b^2)b}{(a^2-b^2)\chi_b}d\psi \, ,
\ee
from which a conformal Killing-Yano tensor can be generated
\be
k = d\hat{b} = r~e^0\wedge e^1 +p~e^2\wedge e^3 \, .
\ee
Adopting the convention $\varepsilon^{01235} = 1 = -\varepsilon_{01235}$ for the totally
antisymmetric tensor density $\varepsilon_{ABCDE}$, we find that the Killing-Yano tensor
is given by
\be
f = {^*}k = \big(-p~e^0\wedge e^1 +r~e^2\wedge e^3\big)\wedge e^5 \, .
\ee
In what follows, we shall show that this rank-three Killing-Yano tensor and its exterior
differential
\be\hspace*{-2cm}
W = df = -4\frac{ab}{rp} ~e^0\wedge e^1\wedge e^2\wedge e^3
 -4\sqrt{\frac{\Delta_p}{\Sigma}} ~e^0\wedge e^1\wedge e^2\wedge e^5
 +4\sqrt{\frac{\Delta_r}{\Sigma}} ~e^1\wedge e^2\wedge e^3\wedge e^5 \, , \quad
\ee
are the essential ingredients to constitute a first order symmetry operator that commutes
with the Dirac operator.

\subsection{First-order symmetry operator from
the separated solution of the Dirac equation}

The final task is to construct a first order symmetry operator that commutes with the Dirac
operator, parallel to the work done in \cite{WuMP} in the case of a five-dimensional Myers-Perry
black hole.

We now proceed to construct such an operator from the separated solutions of the Dirac equation
and highlight the construction procedure. According to our analysis in the last section, we find
that the Dirac equation (\ref{presde}) can be split as
\bea\hspace*{-2cm}
&& \bigg\{ \gamma^0\frac{(r^2+a^2)(r^2+b^2)}{r^2\sqrt{\Delta_r}}\Big(\p_t
 +\frac{a\chi_a}{r^2+a^2}\p_{\phi} +\frac{b\chi_b}{r^2+b^2}\p_{\psi}\Big)
 +\gamma^1\sqrt{\Delta_r}\Big(\p_r
 +\frac{\Delta_r^{\prime}}{4\Delta_r} +\frac{1}{2r}\Big) \nn \\ \hspace*{-2cm}
&&\quad -\frac{i}{r}\big(ab\p_t +b\chi_a\p_{\phi} +a\chi_b\p_{\psi}\big)
 +\frac{iab}{2r^2}\gamma^0\gamma^1 +\mu_er -i\lambda\gamma^0\gamma^1
 \bigg\}\big(\sqrt{r +ip\gamma^5}\Psi\big) = 0 \, , \label{der} \\ \hspace*{-2cm}
&& \bigg\{ \gamma^2\sqrt{\Delta_p}\Big(\p_p +\frac{\Delta_p^{\prime}}{4\Delta_p}
 +\frac{1}{2p}\Big) +\gamma^3\frac{(p^2-a^2)(p^2-b^2)}{p^2\sqrt{\Delta_p}}\Big(\p_t
 -\frac{a\chi_a}{p^2-a^2}\p_{\phi} -\frac{b\chi_b}{p^2-b^2}\p_{\psi}\Big) \nn \\ \hspace*{-2cm}
&&\quad +\frac{\gamma^5}{p}\big(ab\p_t +b\chi_a\p_{\phi} +a\chi_b\p_{\psi}\big)
  +\frac{iab}{2p^2}\gamma^0\gamma^1 -i\mu_ep\gamma^5
  +i\lambda\gamma^0\gamma^1 \bigg\}\big(\sqrt{r +ip\gamma^5}\Psi\big) = 0 \, .
\label{dea}
\eea
Now we multiply Eq. (\ref{der}) by $p\gamma^0\gamma^1$ and Eq. (\ref{dea}) by $-r\gamma^2
\gamma^3$ respectively from the left and then add them together. After using the relations
$i\gamma^5 = \gamma^0\gamma^1\gamma^2\gamma^3$ and $\gamma^2\gamma^3\gamma^5 = i\gamma^0
\gamma^1$, we get a dual equation
\bea\hspace*{-1cm}
&& \bigg\{ \gamma^0p\sqrt{\Delta_r}\Big(\p_r +\frac{\Delta_r^{\prime}}{4\Delta_r}
 +\frac{1}{2r}\Big) +\gamma^1p\frac{(r^2+a^2)(r^2+b^2)}{r^2\sqrt{\Delta_r}}\Big(\p_t
 +\frac{a\chi_a}{r^2+a^2}\p_{\phi} +\frac{b\chi_b}{r^2+b^2}\p_{\psi}\Big) \nn \\ \hspace*{-1cm}
&&\quad +\gamma^2(-r)\frac{(p^2-a^2)(p^2-b^2)}{p^2\sqrt{\Delta_p}}\Big(\p_t
 -\frac{a\chi_a}{p^2-a^2}\p_{\phi} -\frac{b\chi_b}{p^2-b^2}\p_{\psi}\Big) \nn \\ \hspace*{-1cm}
&&\quad +\gamma^3r\sqrt{\Delta_p}\Big(\p_p +\frac{\Delta_p^{\prime}}{4\Delta_p}
  +\frac{1}{2p}\Big) -i\gamma^0\gamma^1\frac{\Sigma}{rp}\big(ab\p_t
 +b\chi_a\p_{\phi} +a\chi_b\p_{\psi}\big) \nn \\ \hspace*{-1cm}
&&\quad +\frac{ab}{2}\Big(\frac{ip}{r^2} +\frac{\gamma^5r}{p^2}\Big)
 +\lambda\big(\gamma^5r -ip\big)\bigg\}\big(\sqrt{r +ip\gamma^5}\Psi\big) = 0 \, .
\eea
Multiplying this equation by the left with a factor $(r+i\gamma^5p)/\Sigma$, we obtain
\bea\hspace*{-1cm}
&& \bigg\{ \gamma^0p\sqrt{\frac{\Delta_r}{\Sigma}}\Big(\p_r
 +\frac{\Delta_r^{\prime}}{4\Delta_r} +\frac{1}{2r} +\frac{r -ip \gamma^5}{2\Sigma}\Big)
 +\gamma^1p\frac{(r^2+a^2)(r^2+b^2)}{r^2\sqrt{\Delta_r\Sigma}}\Big(\p_t
 +\frac{a\chi_a}{r^2+a^2}\p_{\phi} \nn \\ \hspace*{-1cm}
&&\qquad +\frac{b\chi_b}{r^2+b^2}\p_{\psi}\Big)
 +\gamma^2(-r)\frac{(p^2-a^2)(p^2-b^2)}{p^2\sqrt{\Delta_p\Sigma}}\Big(\p_t
 -\frac{a\chi_a}{p^2-a^2}\p_{\phi} -\frac{b\chi_b}{p^2-b^2}\p_{\psi}\Big) \nn \\ \hspace*{-1cm}
&&\qquad +\gamma^3r\sqrt{\frac{\Delta_p}{\Sigma}}\Big[\p_p
 +\frac{\Delta_p^{\prime}}{4\Delta_p} +\frac{1}{2p} +\frac{i\gamma^5}{2\Sigma}(r
 -ip\gamma^5)\Big] -i\gamma^0\gamma^1(r+i\gamma^5p)\frac{1}{rp}\big(ab\p_t \nn \\ \hspace*{-1cm}
&&\qquad\quad +b\chi_a\p_{\phi} +a\chi_b\p_{\psi}\big) +\frac{iab}{2rp}
 +\gamma^5\Big(\lambda -\frac{ab}{2r^2} +\frac{ab}{2p^2}\Big) \bigg\}\Psi = 0 \, .
\eea
We do not hope $\gamma^5\lambda$ appears in the above equation, so we can multiply it
the $\gamma^5$ matrix by the left and rewrite it as
\bea\hspace*{-1.5cm}
&& \bigg\{ \gamma^5\gamma^0p\sqrt{\frac{\Delta_r}{\Sigma}}\Big(\p_r
 +\frac{\Delta_r^{\prime}}{4\Delta_r} +\frac{1}{2r} +\frac{r -ip\gamma^5}{2\Sigma}\Big)
 +\gamma^5\gamma^1p\frac{(r^2+a^2)(r^2+b^2)}{r^2\sqrt{\Delta_r\Sigma}}\Big(\p_t
 +\frac{a\chi_a}{r^2+a^2}\p_{\phi} \nn \\ \hspace*{-1.5cm}
&&\qquad +\frac{b\chi_b}{r^2+b^2}\p_{\psi}\Big)
 +\gamma^5\gamma^2(-r)\frac{(p^2-a^2)(p^2-b^2)}{p^2\sqrt{\Delta_p\Sigma}}\Big(\p_t
 -\frac{a\chi_a}{p^2-a^2}\p_{\phi} -\frac{b\chi_b}{p^2-b^2}\p_{\psi}\Big) \nn \\ \hspace*{-1.5cm}
&&\qquad +\gamma^5\gamma^3r\sqrt{\frac{\Delta_p}{\Sigma}}\Big(\p_p
 +\frac{\Delta_p^{\prime}}{4\Delta_p} +\frac{1}{2p} +\frac{p+i\gamma^5r}{2\Sigma}\Big)
 +\big(p\gamma^0\gamma^1 -r\gamma^2\gamma^3\big)\frac{1}{rp}\big(ab\p_t \nn \\ \hspace*{-1.5cm}
&&\qquad\quad +b\chi_a\p_{\phi} +a\chi_b\p_{\psi}\big) -\frac{ab}{2r^2}
 +\frac{ab}{2p^2} +\frac{iab}{2rp}\gamma^5 +\lambda \bigg\}\Psi = 0 \, .
\eea

The above equation can be viewed as an operator form
\be
\big(\mathbb{H}_f +\lambda\big)\Psi = 0 \, .
\ee
Our final aim is to find the explicit expression for this symmetry operator $\mathbb{H}_f$.
To construct such an operator is more involved than to treat with the Dirac operator
$\mathbb{H}_D = \gamma^{\mu}\nabla_{\mu} = \gamma^{\mu}\big(\p_{\mu} +\Gamma_{\mu}\big)$.

Observing the partial differential terms, we find that it can be exactly given by $-\frac{1}{2}
\gamma^{\mu}\gamma^{\nu}f^{~~\rho}_{\mu\nu}\p_{\rho}$. Therefore, we hope to compute $-\frac{1}{2}
\gamma^{\mu}\gamma^{\nu}f^{~~\rho}_{\mu\nu}\nabla_{\rho}$ in the next step. After some tedious
algebra manipulations, we get its explicit expression as follows:
\bea\hspace*{-1cm}
&& -\frac{1}{2}\gamma^{\mu}\gamma^{\nu}f^{~~\rho}_{\mu\nu} \big(\p_{\rho} +\Gamma_{\rho}\big)
 = \gamma^5\gamma^1p\frac{(r^2+a^2)(r^2+b^2)}{r^2\sqrt{\Delta_r\Sigma}}\Big(\p_t
 +\frac{a\chi_a}{r^2+a^2}\p_{\phi} +\frac{b\chi_b}{r^2+b^2}\p_{\psi}\Big) \nn \\ \hspace*{-1cm}
&&\qquad\quad +\gamma^5\gamma^0p\sqrt{\frac{\Delta_r}{\Sigma}}\Big(\p_r
 +\frac{\Delta_r^{\prime}}{4\Delta_r} +\frac{1}{2r} +\frac{r}{2\Sigma}\Big)
 +\gamma^5\gamma^3r\sqrt{\frac{\Delta_p}{\Sigma}}\Big(\p_p
 +\frac{\Delta_p^{\prime}}{4\Delta_p} +\frac{1}{2p} +\frac{p}{2\Sigma}\Big) \nn \\ \hspace*{-1cm}
&&\qquad\quad  +\gamma^5\gamma^2(-r)\frac{(p^2-a^2)(p^2-b^2)}{p^2\sqrt{\Delta_p\Sigma}}
 \Big(\p_t -\frac{a\chi_a}{p^2-a^2}\p_{\phi}
 -\frac{b\chi_b}{p^2-b^2}\p_{\psi}\Big) \nn \\ \hspace*{-1cm}
&&\qquad\quad +\big(p\gamma^0\gamma^1 -r\gamma^2\gamma^3\big)\frac{1}{rp}
 \big(ab\p_t +b\chi_a\p_{\phi} +a\chi_b\p_{\psi}\big)
 -\frac{ab}{2r^2} +\frac{ab}{2p^2} \nn \\ \hspace*{-1cm}
&&\qquad\quad -\frac{iab}{rp}\gamma^5 +i\sqrt{\frac{\Delta_r}{\Sigma}}\Big(\frac{p^2}{2\Sigma}
 -\frac{3}{2}\Big)\gamma^0 +i\sqrt{\frac{\Delta_p}{\Sigma}}\Big(\frac{3}{2}
 -\frac{r^2}{2\Sigma}\Big)\gamma^3 \, .
\eea
This operator is almost that what we expect to seek for the operator $\mathbb{H}_f$
except the last three terms. These unexpected terms can be remedied by the following
counter-term
\be
-\frac{1}{64}\gamma^{\mu}\gamma^{\nu}\gamma^{\rho}\gamma^{\sigma} W_{\mu\nu\rho\sigma}
 = \frac{3i}{2}\Big(\gamma^0\sqrt{\frac{\Delta_r}{\Sigma}}
 -\gamma^3\sqrt{\frac{\Delta_p}{\Sigma}} +\gamma^5\frac{ab}{rp}\Big) \, .
\ee

Adding them together, we obtain the final expression
\be
\mathbb{H}_f = -\frac{1}{2}\gamma^{\mu}\gamma^{\nu}f^{~~\rho}_{\mu\nu} \big(\p_{\rho}
 +\Gamma_{\rho}\big) -\frac{1}{64}\gamma^{\mu}\gamma^{\nu}\gamma^{\rho}\gamma^{\sigma}
 W_{\mu\nu\rho\sigma} \, .
\ee
Using the definition $W_{\mu\nu\rho\sigma} = -f_{\mu\nu\rho;\sigma} +f_{\nu\rho\sigma;\mu}
-f_{\rho\sigma\mu;\nu} +f_{\sigma\mu\nu;\rho}$ and $f^{\rho}_{~\mu\nu;\rho} = 0$ as well
as the property of gamma matrices, one can also write the above operator in another
equivalent form
\be
\mathbb{H}_f = -\frac{1}{2}\gamma^{\mu}\gamma^{\nu}f^{~~\rho}_{\mu\nu}\nabla_{\rho}
 +\frac{1}{16}\gamma^{\mu}\gamma^{\nu}\gamma^{\rho}\gamma^{\sigma}f_{\mu\nu\rho;\sigma} \, .
\ee

The first order symmetry operator $\mathbb{H}_f$ can be thought of as the ``square root''
of the second order operator $\mathbb{K}$. It has a lot of correspondences in different
contexts. It is a five-dimensional analogue to the nonstandard Dirac operator discovered
by Carter and McLenaghan \cite{CM} for the four-dimensional Kerr metric. This operator
corresponds to the nongeneric supersymmetric generator in pseudo-classical mechanics
\cite{Spcm}. Moreover, the 2-form field $L_{\mu\nu} = f_{\mu\nu\rho} \dot{x}^{\rho}$ is
parallel-propagated along the geodesic with a cotangent vector $\dot{x}^{\mu}$, whose
square is just the Carter's constant $-(1/2)L_{\mu\nu}L^{\mu\nu} = K_{\mu\nu}\dot{x}^{\mu}
\dot{x}^{\nu}$.

The existence of a rank-three Killing-Yano tensor is responsible for the separability of
the Dirac equation in the five-dimensional Kerr-(anti-)de Sitter geometry. The operator
$\mathbb{H}_f$ commutes with the standard Dirac operator $\mathbb{H}_D$. Now we work out
the commutator
\bea\hspace*{-1.5cm}
[\mathbb{H}_f, \mathbb{H}_D] &=& \big[\gamma^{\alpha}\nabla_{\alpha}, \frac{1}{2}
 \gamma^{\mu}\gamma^{\nu}f^{~~\beta}_{\mu\nu}\nabla_{\beta} -\frac{1}{16}\gamma^{\mu}
 \gamma^{\nu}\gamma^{\rho}\gamma^{\sigma}f_{\mu\nu\rho;\sigma}\big] \nn \\ \hspace*{-1.5cm}
&=& \frac{1}{2}\gamma^{\mu}\gamma^{\beta}\gamma^{\nu}f^{~~\alpha}_{\mu\nu}[\nabla_{\alpha},
 \nabla_{\beta}] -\frac{1}{16}\gamma^{\alpha}\gamma^{\mu}\gamma^{\nu}\gamma^{\rho}
 \gamma^{\sigma}f_{\mu\nu\rho;\sigma;\alpha} \nn \\ \hspace*{-1.5cm}
&& +\frac{1}{2}\gamma^{\alpha}\gamma^{\mu}\gamma^{\nu}\nabla_{\alpha}f^{~~\beta}_{\mu\nu}
 \nabla_{\beta} +\frac{1}{16}[\gamma^{\mu}\gamma^{\nu}\gamma^{\rho}\gamma^{\sigma},
 \gamma^{\beta}]f_{\mu\nu\rho;\sigma}\nabla_{\beta} \nn \\ \hspace*{-1.5cm}
&=& \frac{1}{8}\gamma^{\mu}\gamma^{\beta}\gamma^{\nu}\gamma^{\rho}\gamma^{\sigma}
 f^{~~\alpha}_{\mu\nu}R_{\alpha\beta\rho\sigma} -\frac{1}{16}\gamma^{\alpha}\gamma^{\mu}
 \gamma^{\nu}\gamma^{\rho}\gamma^{\sigma}f_{\mu\nu\rho;\sigma;\alpha} \nn \\
&& +\frac{1}{8}\gamma^{\alpha}\gamma^{\mu}\gamma^{\nu}\nabla_{\alpha}f_{\mu\nu}^{~~\beta}
 \nabla_{\beta} +\frac{1}{8}\gamma^{\mu}\gamma^{\nu}\gamma^{\alpha}f_{\mu\nu\alpha;\beta}
 \nabla^{\beta} \nn \\ \hspace*{-1.5cm}
&=& \frac{1}{16}\gamma^{\alpha}\gamma^{\mu}\gamma^{\nu}\gamma^{\rho}\gamma^{\sigma}
 \big(2f^{~~\beta}_{\alpha\nu}R_{\beta\mu\rho\sigma} -f_{\mu\nu\rho;\sigma;\alpha}\big)
  +\frac{1}{8}\gamma^{\alpha}\gamma^{\mu}\gamma^{\nu}\big(f_{\mu\nu\alpha;\beta}
 +f_{\mu\nu\beta;\alpha}\big)\nabla^{\beta} \, .
\eea
To derive the last expression for the commutator, we have made use of the anti-commutativity
of Dirac gamma matrices and the following relations
\be
\nabla_{\mu}\gamma^{\nu} = 0 \, , \qquad [\nabla_{\mu}, \nabla_{\nu}] =
 \frac{1}{4}R_{\mu\nu\rho\sigma}\gamma^{\rho}\gamma^{\sigma} \, , \qquad
 f^{\rho}_{~\mu\nu;\rho} = 0 \, .
\ee

Then we can see that the commutation relation $[\mathbb{H}_f, \mathbb{H}_D] = 0$ just yields
the Killing-Yano equation (\ref{KYe}) and the integrability condition for the rank-three
Killing-Yano tensor.

\section{Conclusions}
\label{CoRe}

In this paper, we have investigated the separability of the Dirac equation in the five-dimensional
Kerr-(anti-)de Sitter metric and its relation to a rank-three Killing-Yano tensor. We work out
the Dirac field equation within a f\"{u}nfbein formalism. The spinor connection is constructed
by the method of the Clifford algebra and its derived Lie algebra SO(4,1). We establish a local
orthonormal pentad for the Kerr-(anti-)de Sitter metric so that we can easily copy with the
Dirac equation in this background geometry. We then obviously show that Dirac's equation in
the Kerr-(anti-)de Sitter metric can be separated into purely radial and purely angular parts.
From the separated solutions of the massive Klein-Gordon equation and Dirac's equation, we have
constructed two symmetry operators that commute with the scalar Laplacian operator and the Dirac
operator, respectively. A simple form for the St\"{a}ckel-Killing tensor was given so that it
can be easily understood as the square of a rank-three Killing-Yano tensor.

Our work in this paper includes the previous one \cite{WuMP} as a special case done for the
general $D = 5$ Myers-Perry metric. The results presented here can serve as a basis to study
various aspects \cite{App} of the Dirac field, such as Hawking radiation, quasinormal modes,
instability, supersymmetry, etc. In our forthcoming papers \cite{WuNP}, the present work has
been extended to the charged case of $D = 5$ rotating black holes in minimal gauged \cite{EMCS}
and ungauged supergravity \cite{CYL} with the inclusion of a Chern-Simons term. It is found
that the usual Dirac equation can not be separated by variables. With the supplement of an
additional term into the equation of fermion fields, it is shown that the modified Dirac
equation can be decoupled into purely radial and purely angular parts in these
Einstein-Maxwell-Chern-Simons background spacetimes. A paper about this aspect is
in preparation.

It is an open question to investigate the separability of higher-spin field equations (for
example, Maxwell's equation and Rarita-Schwinger's equation) in the five-dimensional
Kerr-(anti-)de Sitter metric and its relation to a rank-three Killing-Yano tensor.

\section*{Acknowledgements}

This work is partially supported by the Natural Science Foundation of China under Grant No.
10675051.

\section*{Appendix:~~ Pentad, connection one-forms,
and curvature two-forms}

\def\theequation{A\arabic{equation}}
\setcounter{equation}{0}

The new form of the five-dimensional Kerr-(anti-)de Sitter metric (\ref{KMP}) admits the
following local Lorentz basis of one-forms (pentad) defined as $e^A = e^A_{~\mu}dx^{\mu}$
orthonormal with respect to $\eta_{AB}$,
\be\hspace*{-1.5cm}
 e^0 = \sqrt{\frac{\Delta_r}{\Sigma}} X \, , \quad
 e^1 = \sqrt{\frac{\Sigma}{\Delta_r}}dr \, , \quad
 e^2 = \sqrt{\frac{\Sigma}{\Delta_p}}dp \, , \quad
 e^3 = \sqrt{\frac{\Delta_p}{\Sigma}} Y \, , \quad
 e^5 = -\frac{ab}{rp} Z \, .
\label{pentad}
\ee
The above pentad is different from those used in \cite{ANA}. The Dirac equation is difficult
to be decoupled if one adopts the pentad in \cite{ANA}.

After some algebraic computations, we obtain the exterior differential of the coframe one-forms
\numparts\bea
&& de^0 = -\Big(\frac{\Delta_r^{\prime}}{2\Delta_r} -\frac{r}{\Sigma}\Big)
 \sqrt{\frac{\Delta_r}{\Sigma}} ~e^0\wedge e^1
 -\frac{p}{\Sigma}\sqrt{\frac{\Delta_p}{\Sigma}} ~e^0\wedge e^2
 -\frac{2p}{\Sigma} \sqrt{\frac{\Delta_r}{\Sigma}} ~e^2\wedge e^3 \, ,  \\
&& de^1 = -\frac{p}{\Sigma}\sqrt{\frac{\Delta_p}{\Sigma}} ~e^1\wedge e^2 \, , \\
&& de^2 = \frac{r}{\Sigma}\sqrt{\frac{\Delta_r}{\Sigma}} ~e^1\wedge e^2 \, , \\
&& de^3 = \frac{2r}{\Sigma}\sqrt{\frac{\Delta_p}{\Sigma}} ~e^0\wedge e^1
 +\frac{r}{\Sigma}\sqrt{\frac{\Delta_r}{\Sigma}} ~e^1\wedge e^3
 +\Big(\frac{\Delta_p^{\prime}}{2\Delta_p} -\frac{p}{\Sigma}\Big)
 \sqrt{\frac{\Delta_p}{\Sigma}} ~e^2\wedge e^3 \, , \\
&& de^5 = -\frac{2ab}{r^2p} ~e^0\wedge e^1
 +\frac{1}{r}\sqrt{\frac{\Delta_r}{\Sigma}} ~e^1\wedge e^5
 +\frac{2ab}{rp^2} ~e^2\wedge e^3 +\frac{1}{p}
 \sqrt{\frac{\Delta_p}{\Sigma}} ~e^2\wedge e^5 \, .
\eea\endnumparts

The spin-connection one-form $\omega^A_{~B} = \omega^A_{~B\mu}dx^{\mu} = f^A_{~BC}e^C$ can be
found from the Cartan's first structure equation (\ref{CFE}) as follows:
\bea
&& \omega^0_{~1} = \Big(\frac{\Delta_r^{\prime}}{2\Delta_r} -\frac{r}{\Sigma}\Big)
 \sqrt{\frac{\Delta_r}{\Sigma}} ~e^0 +\frac{r}{\Sigma}\sqrt{\frac{\Delta_p}{\Sigma}} ~e^3
  -\frac{ab}{r^2p} ~e^5 \, , \nn \\
&& \omega^0_{~2} = \frac{p}{\Sigma}\sqrt{\frac{\Delta_p}{\Sigma}} ~e^0
 -\frac{p}{\Sigma}\sqrt{\frac{\Delta_r}{\Sigma}} ~e^3 \, , \nn \\
&& \omega^0_{~3} = \frac{r}{\Sigma}\sqrt{\frac{\Delta_p}{\Sigma}} ~e^1
 +\frac{p}{\Sigma}\sqrt{\frac{\Delta_r}{\Sigma}} ~e^2 \, , \nn \\
&& \omega^0_{~5} =  -\frac{ab}{r^2p} ~e^1 \, , \nn \\
&& \omega^1_{~2} = \frac{p}{\Sigma}\sqrt{\frac{\Delta_p}{\Sigma}} ~e^1
 -\frac{r}{\Sigma}\sqrt{\frac{\Delta_r}{\Sigma}} ~e^2 \, , \nn \\
&& \omega^1_{~3} = \frac{r}{\Sigma}\sqrt{\frac{\Delta_p}{\Sigma}} ~e^0
 -\frac{r}{\Sigma}\sqrt{\frac{\Delta_r}{\Sigma}} ~e^3 \, , \nn \\
&& \omega^1_{~5} =  -\frac{ab}{r^2p} ~e^0
 -\frac{1}{r}\sqrt{\frac{\Delta_r}{\Sigma}} ~e^5 \, , \nn \\
&& \omega^2_{~3} = -\frac{p}{\Sigma}\sqrt{\frac{\Delta_r}{\Sigma}} ~e^0
 -\Big(\frac{\Delta_p^{\prime}}{2\Delta_p} -\frac{p}{\Sigma}\Big)
 \sqrt{\frac{\Delta_p}{\Sigma}} ~e^3 -\frac{ab}{rp^2} ~e^5 \, , \nn \\
&& \omega^2_{~5} = -\frac{ab}{rp^2} ~e^3
 -\frac{1}{p}\sqrt{\frac{\Delta_p}{\Sigma}} ~e^5 \, , \nn \\
&& \omega^3_{~5} = \frac{ab}{rp^2} ~e^2 \, .
\eea
The local Lorentz frame component $\Gamma_A$ can be easily read from the spinor connection
one-form $\Gamma \equiv \Gamma_Ae^A = (1/4)\gamma^A\gamma^B\omega_{AB}$ and is given by
\bea
&& \Gamma_0 = -\Big(\frac{\Delta_r^{\prime}}{4\Delta_r}
 -\frac{r}{2\Sigma}\Big)\sqrt{\frac{\Delta_r}{\Sigma}}\gamma^0\gamma^1
 -\frac{p}{2\Sigma}\sqrt{\frac{\Delta_p}{\Sigma}}\gamma^0\gamma^2 \nn \\
&&\qquad\quad +\frac{r}{2\Sigma}\sqrt{\frac{\Delta_p}{\Sigma}}\gamma^1\gamma^3
 -\frac{ab}{2r^2p}\gamma^1\gamma^5
 -\frac{p}{2\Sigma}\sqrt{\frac{\Delta_r}{\Sigma}}\gamma^2\gamma^3 \, , \nn \\
&& \Gamma_1 = -\frac{r}{2\Sigma}\sqrt{\frac{\Delta_p}{\Sigma}}\gamma^0\gamma^3
 +\frac{ab}{2r^2p}\gamma^0\gamma^5
 +\frac{p}{2\Sigma}\sqrt{\frac{\Delta_p}{\Sigma}}\gamma^1\gamma^2 \, , \nn \\
&& \Gamma_2 = -\frac{p}{2\Sigma}\sqrt{\frac{\Delta_r}{\Sigma}}\gamma^0\gamma^3
 -\frac{r}{2\Sigma}\sqrt{\frac{\Delta_r}{\Sigma}}\gamma^1\gamma^2
 +\frac{ab}{2rp^2}\gamma^3\gamma^5 \, , \nn \\
&& \Gamma_3 = -\frac{r}{2\Sigma}\sqrt{\frac{\Delta_p}{\Sigma}}\gamma^0\gamma^1
 +\frac{p}{2\Sigma}\sqrt{\frac{\Delta_r}{\Sigma}}\gamma^0\gamma^2
 -\frac{r}{2\Sigma}\sqrt{\frac{\Delta_r}{\Sigma}}\gamma^1\gamma^3 \nn \\
&&\qquad\quad -\Big(\frac{\Delta_p^{\prime}}{4\Delta_p} -\frac{p}{2\Sigma}\Big)
 \sqrt{\frac{\Delta_p}{\Sigma}}\gamma^2\gamma^3 -\frac{ab}{rp^2}\gamma^2\gamma^5 \, , \nn \\
&& \Gamma_5 = \frac{ab}{2r^2p}\gamma^0\gamma^1
 -\frac{1}{2r}\sqrt{\frac{\Delta_r}{\Sigma}}\gamma^1\gamma^5
 -\frac{ab}{2rp^2}\gamma^2\gamma^3
 -\frac{1}{2p}\sqrt{\frac{\Delta_p}{\Sigma}}\gamma^2\gamma^5 \, .
\label{Lsc}
\eea

Using the above pentad formalism, the curvature two-forms $\mathcal{R}^A_{~B} = d\omega^A_{~B}
+\omega^A_{~C}\wedge \omega^C_{~B}$ can be concisely expressed by
\bea
&& \mathcal{R}^0_{~1} = \alpha ~e^0\wedge e^1 -2C_0 ~e^2\wedge e^3 \, , \nn \\
&& \mathcal{R}^0_{~2} = \beta ~e^0\wedge e^2 -C_0 ~e^1\wedge e^3 \, , \nn \\
&& \mathcal{R}^0_{~3} = \beta ~e^0\wedge e^3 +C_0 ~e^1\wedge e^2 \, , \nn \\
&& \mathcal{R}^0_{~5} = \gamma ~e^0\wedge e^5 \, , \nn \\
&& \mathcal{R}^1_{~2} = -C_0 ~e^0\wedge e^3 +\beta ~e^1\wedge e^2 \, , \nn \\
&& \mathcal{R}^1_{~3} = C_0 ~e^0\wedge e^2 +\beta ~e^1\wedge e^3 \, , \nn \\
&& \mathcal{R}^1_{~5} = \gamma ~e^1\wedge e^5 \, , \nn \\
&& \mathcal{R}^2_{~3} = 2C_0 ~e^0\wedge e^1 +\delta ~e^2\wedge e^3 \, , \nn \\
&& \mathcal{R}^2_{~5} = \tilde{\epsilon} ~e^2\wedge e^5 \, , \nn \\
&& \mathcal{R}^3_{~5} = \tilde{\epsilon} ~e^3\wedge e^5 \, ,
\eea
where
\bea\hspace*{-1cm}
&& \alpha = \frac{\epsilon}{l^2} +\frac{2M}{\Sigma^3}\big(3r^2-p^2\big) \, ,  \qquad
  \beta  = \frac{\epsilon}{l^2} -\frac{2M}{\Sigma^3}\big(r^2-p^2\big) \, , \qquad
  \gamma = \frac{\epsilon}{l^2} -\frac{2M}{\Sigma^2} \, , \nn \\ \hspace*{-1cm}
&&\quad \delta = \frac{\epsilon}{l^2} +\frac{2M}{\Sigma^3}\big(r^2-3p^2\big) \, , \qquad
  \tilde{\epsilon} = \frac{\epsilon}{l^2} +\frac{2M}{\Sigma^2} \, , \qquad\quad
  C_0 = \frac{4Mrp}{\Sigma^3} \, .
\eea
Finally, Ricci tensors and the scalar curvature for the $D = 5$ Kerr-(anti-)de Sitter metric
are
\be
R_{AB} = \eta_{AB}\frac{4\epsilon}{l^2} \, , \qquad R = \frac{20\epsilon}{l^2} \, .
\ee

\section*{References}

\providecommand{\href}[2]{#2}\begingroup\raggedright

\endgroup

\end{document}